\documentclass{article}
\usepackage{spconf,amsmath,graphicx}
\usepackage{amsmath,amssymb,amsfonts}
\usepackage{algorithmic}
\usepackage{hyperref}
\usepackage{pifont}
\usepackage{xcolor}
\usepackage{booktabs}
\usepackage{threeparttable}
\usepackage{multirow}
\usepackage{makecell}
\title{MELA-TTS: JOINT TRANSFORMER-DIFFUSION MODEL WITH REPRESENTATION ALIGNMENT FOR SPEECH SYNTHESIS}
\name{\parbox{\linewidth}{\centering Keyu An$^{\star}$, Zhiyu Zhang$^{\star \dagger}$, Changfeng Gao$^{\star}$, Yabin Li$^{\star}$,  Zhendong Peng$^{\star}$, \\ Haoxu Wang$^{\star}$, Zhihao Du$^{\star}$, Han Zhao$^{\star}$, Zhifu Gao$^{\star}$, Xiangang Li$^{\star}$ \thanks{The first two authors contribute equally to this work.}}}
\address{$^{\star}$ Alibaba group 
$^{\dagger}$ National Mobile Communications Research Laboratory, Southeast University \\
ankeyu.aky@alibaba-inc.com, zhiyuzhang@seu.edu.cn}
\begin{document}
\ninept
\maketitle
\begin{abstract}
This work introduces MELA-TTS, a novel joint transformer-diffusion framework for end-to-end text-to-speech synthesis. By autoregressively generating continuous mel-spectrogram frames from linguistic and speaker conditions, our architecture eliminates the need for speech tokenization and multi-stage processing pipelines. To address the inherent difficulties of modeling continuous features, we propose a representation alignment module that aligns output representations of the transformer decoder with semantic embeddings from a pretrained ASR encoder during training. This mechanism not only speeds up training convergence, but also enhances cross-modal coherence between the textual and acoustic domains. Comprehensive experiments demonstrate that MELA-TTS achieves state-of-the-art performance across multiple evaluation metrics while maintaining robust zero-shot voice cloning capabilities, in both offline and streaming synthesis modes. Our results establish a new benchmark for continuous feature generation approaches in TTS, offering a compelling alternative to discrete-token-based paradigms.
\end{abstract}
\section{Introduction}
\label{sec:intro}
Autoregressive modeling based on discrete tokens has demonstrated remarkable success in text-to-speech (TTS) synthesis. Such frameworks critically depend on a pre-trained tokenizer to discretize continuous speech features into token sequences~\cite{valle,cosyvoice3}. During the generation process, an autoregressive model first performs next-token prediction, after which a dedicated decoder network maps the discrete tokens back to high-dimensional continuous speech features. While demonstrating exceptional proficiency in achieving high-fidelity speech naturalness and cross-speaker generalization through zero-shot voice cloning, this framework exhibits inherent limitations. First, the discretization of speech signals inherently incurs information loss, which fundamentally constrains the fidelity of subsequent speech reconstruction. Second and critically, the decoupled two-stage framework increases system complexity while creating a cascading error accumulation. 

Recent studies have proposed end-to-end frameworks that directly generate continuous speech features without relying on discrete token intermediates~\cite{melle, ditar}. This paradigm shift eliminates the need for multi-stage pipelines while preserving the full information of raw speech features. However, these architectures still face critical challenges. Firstly, their performance lags behind state-of-the-art discrete-token-based models, particularly in content consistency~\cite{kalle}. Recently proposed DiTAR~\cite{ditar} achieves remarkable results in terms of low WER/CER on benchmarks. However, it's not clear whether it's robust on hard cases, such as long text containing repetitions, tongue twisters, and so on. Secondly, end-to-end architectures introduce significant optimization challenges: autoregressive modeling of continuous features typically requires substantially more training iterations to converge, compared to discrete-token-based autoregressive frameworks, due to the inherent complications in modeling high-dimensional continuous features. 
% In addition, to meet the low-latency requirements of real-time applications, multiple streaming zero-shot TTS systems have been developed\cite{cosyvoice2,sheng2025syncspeech,smlle}, which inevitably yield suboptimal reconstruction quality due to the reliance on discrete representations.

To address these challenges, we propose MELA-TTS\footnote{MELA is the combination of MEL and Alignment.}, a joint transformer and diffusion model that generates mel-spectrogram auto-regressively, eliminating the need for a speech tokenizer and multi-stage training and inference pipelines. To enhance content consistency and facilitate training convergence, we propose a representation alignment module that aligns the model's intermediate representations with those extracted from a pre-trained ASR encoder. The efficacy of the proposed method is validated through comprehensive experiments in two scenarios: (1) {\bf offline synthesis}, which requires a complete text as input, and (2) {\bf streaming synthesis}, where the input text is received in a streaming manner rather than given as a complete sentence in advance. Moreover, the scalability of MELA-TTS is evidenced by remarkable performance improvement when the training data scales to 170,000 hours.
% Furthermore, we interleave the continuous text embeddings and mel representations at an $n:m$ ratio that enables incremental TTS synthesis while mitigating information loss from quantization.  Unlike CosyVoice2\cite{cosyvoice2} which necessitates buffering $m$ speech tokens, MELA-TTS can synthesize mel-spectrogram chunks per transformer hidden state through a chunk-aware diffusion model.

%The main contributions are summarized as follows:

\begin{figure}[!t]

  \centering
  \centerline{\includegraphics[width=7cm]{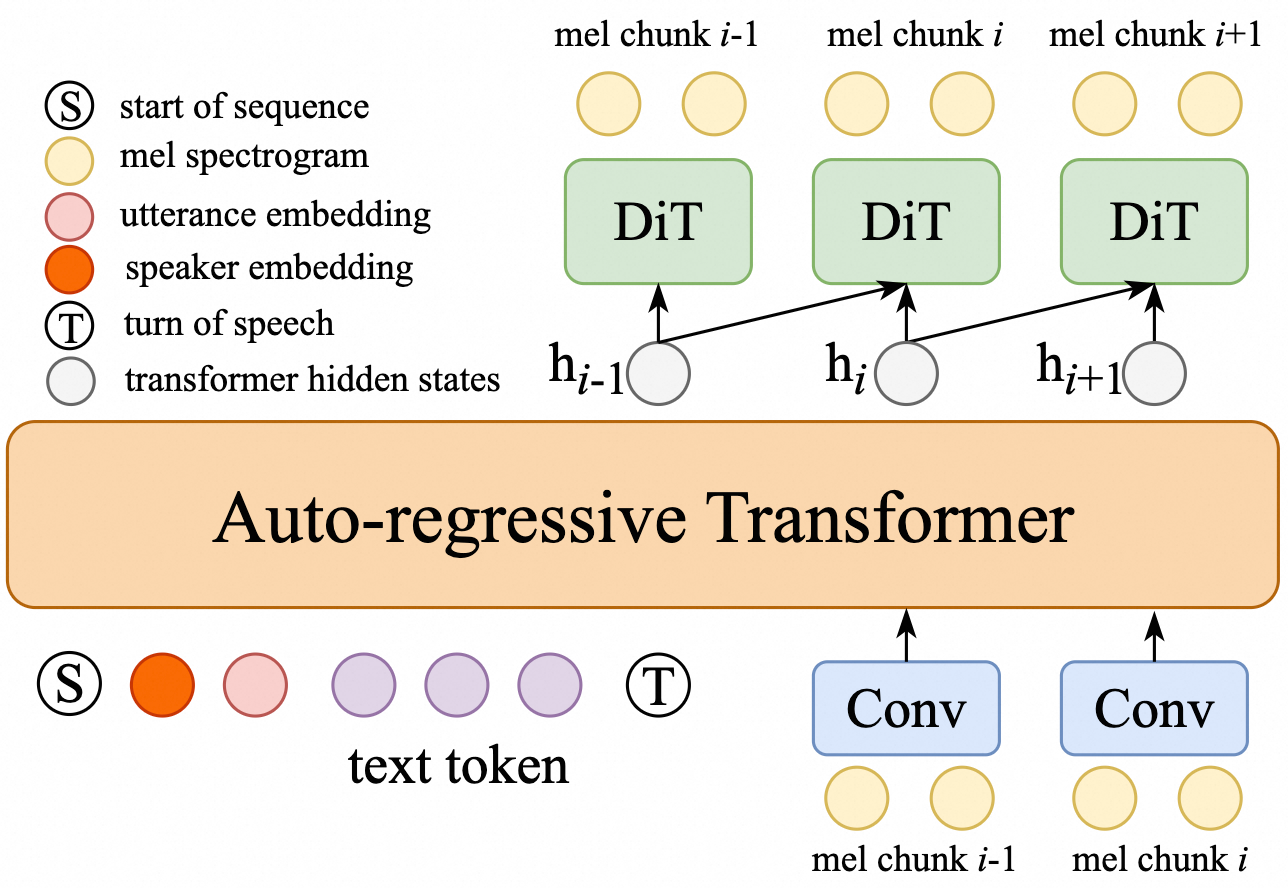}}
  \vspace{-0.5cm}
%  \vspace{2.0cm}
  % \centerline{}\medskip
\caption{The joint transformer and diffusion architecture. The autoregressive transformer decoder generates continuous vectors $\bf{h}$ as the condition to the diffusion model to generate the mel chunks.}
\label{fig:model}
\vspace{-0.5cm}
\end{figure}

\section{Methods}
\label{sec:method}
As illustrated in the Figure~\ref{fig:model}, MELA-TTS comprises an autoregressive transformer decoder and a diffusion module. The autoregressive transformer decoder generates continuous vectors $\bf{h}$ sequentially, and the diffusion module utilizes these vectors, along with speaker embeddings and utterance embeddings as conditional inputs, to perform a denoising process on the noisy mel-spectrogram chunk. Once the mel-spectrogram is generated, the speech waveform can be constructed using a neural vocoder. More importantly, we introduced a representation alignment module to align the continuous vectors $\bf{h}$ with the output representations of a pretrained ASR encoder, which encourages $\bf{h}$ to be more semantically informative, thereby improving the content consistency of the generated outputs. The detailed descriptions of each module are presented below.

\subsection{Transformer decoder for auto-regressive modeling}
\label{sec:transformer}
In MELA-TTS, a transformer decoder autoregressively generates continuous vectors $\bf{h}$ conditioned on the utterance embedding, the speaker embedding, the tokenized text, and the mel-spectrogram history ${\bf X}=[x_1, x_2, ..., x_L]$. During training, the utterance embedding is extracted from randomly cropped segments of the input speech via a transformer encoder. The transformer encoder outputs features that are pooled into an utterance embedding vector, and is jointly optimized with the transformer decoder and the diffusion model.  The speaker embedding is captured from the input speech with a pretrained speaker encoder~\footnote{https://github.com/modelscope/3D-Speaker}. During inference, both the utterance embedding and speaker embedding are derived from the prompt speech. The text input is first tokenized into BPE tokens with the tokenizer from Qwen2, and then converted to embeddings using Qwen 2's text embedding layer. As for the mel-spectrogram history, the $i\mbox{-}\rm{th}$ chunk of mel-spectrogram ${\bf X}^{(i)}=[x_{i\times N + 1}, ..., x_{(i+1)\times N}] \in \mathcal{R}^{N \times D_{\rm mel}}$ is downsampled and projected into a tensor of shape $[1, D_{\rm trans}]$ by a strided convolution layer, and then fed into the transformer decoder.  Here $N$ is the chunk size, $D_{\rm mel}$ is the dimension of the mel-spectrogram, and $D_{\rm trans}$ is the dimension of the transformer decoder. Following~\cite{kaiming_ardiff}, the output of the final transformer decoder layer $\bf{h}$ will serve as the condition for the diffusion model.

Unlike discrete-token-based TTS systems that terminate generation via prediction of a special end-of-sequence (EOS) token, MELA-TTS employs a stop prediction module to determine the end of synthesize. This module functions as a binary classifier: it takes the continuous hidden representation sequence $\bf{h}$ as input and outputs a binary decision (0/1) at each step, where 0 signifies continuation and 1 indicates termination of the speech synthesis process. The module is trained using a binary cross-entropy (BCE) loss $\mathcal{L}_{\rm stop}$.

\begin{figure}[!t]

  \centering
  \centerline{\includegraphics[width=7.5cm]{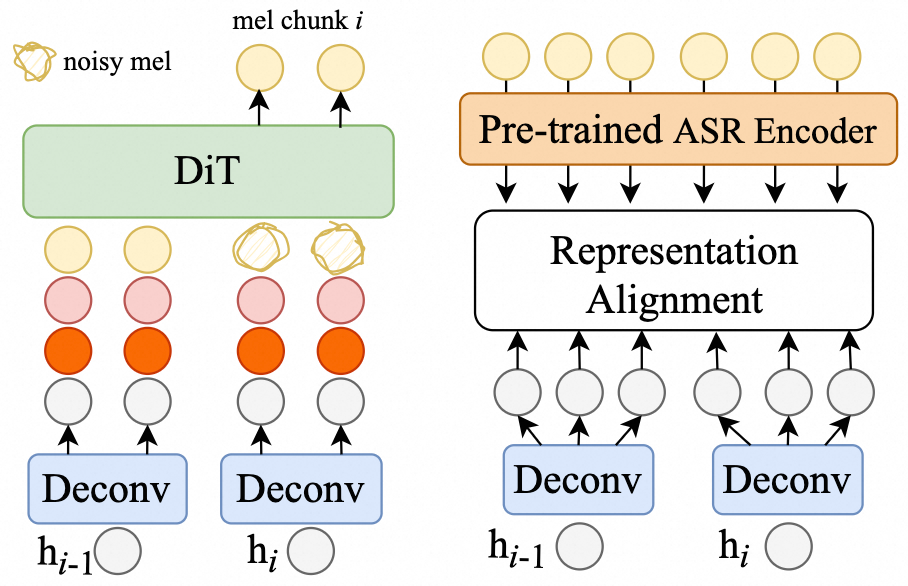}}
  \vspace{-0.5cm}
  % \centerline{}\medskip
\caption{Left: the diffusion module utilizes $\bf{h}$, along with speaker embeddings $\bf{v}$ and utterance embeddings $\bf{u}$ as conditional inputs, to perform mel-spectrogram denoising. $\bf{h}$, $\bf{v}$, and $\bf{u}$ are upsampled respectively to align with the chunk size of the mel-spectrogram. Right: the representation alignment module. $\bf{h}$ is also upsampled to align with the length of the pretrained semantic representation.}
\label{fig:diff_align}
\vspace{-0.5cm}
\end{figure}

\subsection{Diffusion for mel-spectrogram generation}
\label{sec:diffusion}
In MELA-TTS, the diffusion module, implemented as a diffusion transformer~\cite{dit}, predicts a chunk of mel-spectrogram ${\bf X}^{(i)}_0 := {\bf X}^{(i)}$ based on $[h_{i-1}, h_i]$, speaker embeddings $\bf{v}$, utterance embeddings $\bf{u}$, and noisy mel-spectrogram chunk with previous mel-spectrogram chunk prepended $[{\bf X}^{(i-1)}_0, {\bf X}^{(i)}_t]$:
\begin{equation}
\begin{aligned}
    \hat{{\bf X}}^{(i)}_0  &= {\rm DiT}(\Psi_i,[{\bf X}^{(i-1)}_0, {\bf X}^{(i)}_t]) \\ 
    &= {\rm DiT}([h_{i-1}, h_i], {\bf v}, {\bf u}, [{\bf X}^{(i-1)}_0, {\bf X}^{(i)}_t]).
    \nonumber
\end{aligned}
\end{equation}
Here $\Psi_i=\{[h_{i-1}, h_i], {\bf v}, {\bf u}\}$ is the condition, and ${\bf X}^{(i)}_t= \alpha_t {\bf X}^{(i)}_0 + \sigma_t \epsilon$ is given by a diffusion forward process~\cite{song2020score}, and $\epsilon$ is the standard gaussian noise. We follow the variance preserving (VP) formulation and set $\alpha_t=\rm{cos}(\frac{\pi t}{2})$ and $\sigma_t=\rm{sin}(\frac{\pi t}{2})$. The previous continuous vector $h_{i-1}$ and the mel-spectrogram chunk ${\bf X}^{(i-1)}_0$ are provided as prefix context for the diffusion model, and the output of the prefix part will be discarded. The loss is defined as the L2 distance between the prediction and the ground truth mel-spectrogram:
$$ \mathcal{L}_{\rm diff} = \sum_{i}(\hat{{\bf X}}^{(i)}_0 - {\bf X}^{(i)}_0)^2 .$$ 

\subsection{Representation alignment module}
\label{sec:align}
In discrete-token-based TTS systems, supervised semantic tokens, which are typically derived from an ASR model, have demonstrated superior efficacy as intermediate representations, significantly improving content consistency and voice cloning performance~\cite{cosyvoice}. However, in end-to-end models, since the model directly predicts mel-spectrograms or other continuous representations, it is not explicitly guided to produce semantically enriched intermediates. The absence of intermediate semantic guidance leads to two adverse consequences: poor content consistency in the synthesized speech, and slower convergence during model training. To address it, we propose a representation alignment module, as illustrated in Figure~\ref{fig:diff_align}. Specifically, we align the output of the autoregressive transformer $\bf{h}$ with pretrained semantic representations $\bf{h_{asr}}$ generated by an ASR encoder by adding a cosine similarity loss term between them:
$$ \mathcal{L}_{\rm align} = {\rm CosineSimilarity}({\rm TAM}({\bf h}), {\bf h_{asr}}), $$
where ${\rm TAM}$ is a time alignment module to resolve temporal resolution mismatches between ${\bf h}$ and ${\bf h_{asr}}$, implemented as a linear layer followed by reshape operations.

For alignment objective, one might intuitively consider using mel-spectrogram directly as the alignment target. However, experimental results revealed that this strategy fails to provide positive gains and instead significantly degrades both content consistency and speaker similarity in voice cloning. We will discuss it in~\ref{sec:ablation}.

To sum up, the overall training loss is defined as:
$$ \mathcal{L} = \mathcal{L}_{\rm diff} + \mathcal{L}_{\rm stop} + \mathcal{L}_{\rm align} $$
\subsection{Streaming synthesis}
\label{sec:streaming}
\begin{figure}[!t]

  \centering
  \centerline{\includegraphics[width=8cm]{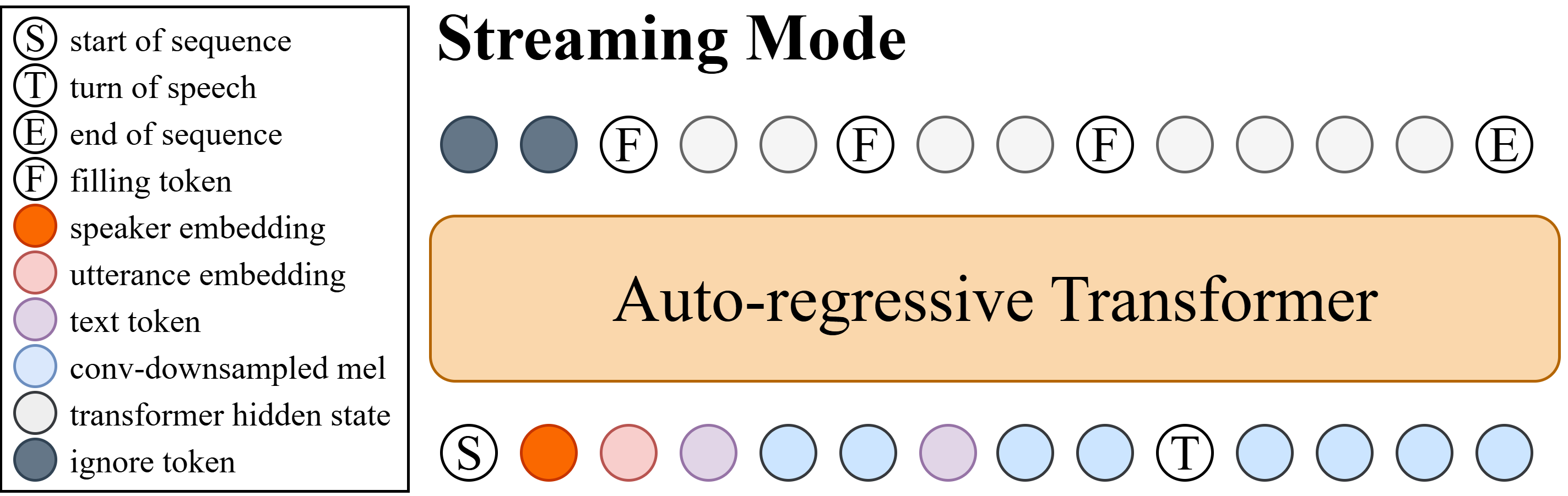}}
  \vspace{-0.45cm}
%  \vspace{2.0cm}
  % \centerline{}\medskip
\caption{A diagram of the auto-regressive language model for streaming synthesis in MELA-TTS.}
\label{fig:stream}
\vspace{-0.45cm}
\end{figure}

For streaming synthesis (Figure~\ref{fig:stream}), we interleave the text tokens and continuous conv-downsampled mel-spectrogram in an $n:m$ ratio, which enables incremental speech synthesis and allows the generation of $m$ mel-spectrogram chunks for every $n$ text tokens received. The model is simultaneously trained on both interleaved and non-interleaved sequences, thus streaming and non-streaming synthesis can be performed within a unified model. The turn-of-speech token indicates the end of text input, and the filling token only marks the position and is excluded for target prediction and loss calculation. The termination of speech generation is determined by the binary classification module, which is the same as the offline model.

% Since no post-processing module is adopted in MELA-TTS, the streaming mode of TTS synthesize can reduce the fisrt-package latency from $L=N\cdot d_{LLM}+M\cdot d_{TTS}$ to $L=n\cdot d_{LLM}+d_{TTS}$, where $d_{LLM}$ represents the time required by the LLM to generate one text token, $d_{TTS}$ denotes the waveform reconstruction delay per mel-spectrogram chunk, and $N$ and $M$ indicate the total counts of text tokens and mel-spectrogram chunks, respectively.

\begin{figure}[!t]

  \centering
  \centerline{\includegraphics[width=6cm]{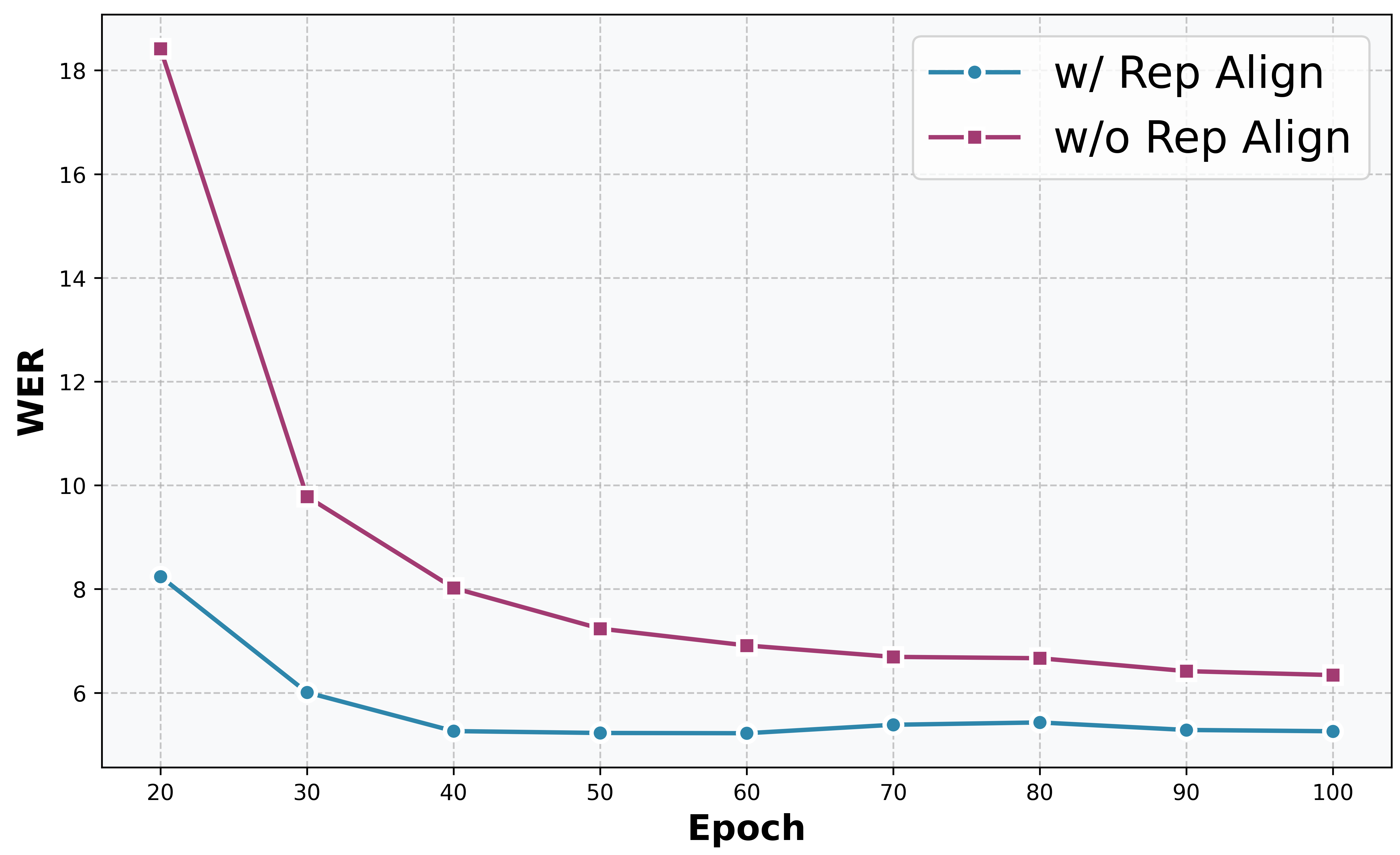}}
  \vspace{-0.5cm}
%  \vspace{2.0cm}
  % \centerline{}\medskip
\caption{Comparison of WER over training epochs with and without representation alignment.}
% The blue line (w/ Rep Align) shows consistently lower WER and faster convergence compared to the magenta line (w/o Rep Align)
\label{fig:wer_convergence}
\vspace{-0.5cm}
\end{figure}

\begin{table}
    \centering
    %\huge
    \vspace{-0.8cm}
    \caption{Ablation study of streaming synthesis, utterance embedding (Utt Emb), and representation alignment (Rep Align). $\ast$ indicates using mel-spectrogram instead of the pretrained ASR encoder output as the representation alignment target.}
    % SS1 is calculated using WavLM-based speaker verification model, and SS1 is calculated using ERes2Net-based speaker verification model.
    \resizebox{0.4\textwidth}{!}{
    {
    \begin{tabular}{cccc|cccc}
        \toprule
         \textbf{Exp} & \textbf{Strea} & \textbf{Utt}&  \textbf{Rep}&   \multirow{2}*{WER~$\downarrow$}&  \multirow{2}*{SS1~$\uparrow$}& \multirow{2}*{SS2~$\uparrow$} \\
         \textbf{ID} & \textbf{ming} & \textbf{Emb} &  \textbf{Align}&   &  & \\         
         
         %\textbf{Streaming} & \textbf{Utt Emb} & \textbf{Rep Align} & \textbf{WER~$\downarrow$} & \textbf{SS1~$\uparrow$} & \textbf{SS2~$\uparrow$} \\
        \midrule
     0& \textcolor{red}{\ding{55}}& \textcolor{red}{\ding{55}}& \textcolor{red}{\ding{55}}& 6.3& 0.46 & 0.55\\
     1&\textcolor{red}{\ding{55}}& \textcolor{red}{\ding{55}}& \textcolor{green}{\ding{51}}& 5.3& 0.46 &0.54 \\
     2&\textcolor{red}{\ding{55}}& \textcolor{red}{\ding{55}}& \textcolor{green}{\ding{51}} $\ast$ & 6.7& 0.41 &0.48 \\
     3&\textcolor{red}{\ding{55}}& \textcolor{green}{\ding{51}}& \textcolor{red}{\ding{55}}& 6.0&0.47 &0.57 \\
     4&\textcolor{red}{\ding{55}}& \textcolor{green}{\ding{51}}& \textcolor{green}{\ding{51}}& 5.2& \textbf{0.48} & \textbf{0.58} \\
     \midrule
     5&\textcolor{green}{\ding{51}}& \textcolor{red}{\ding{55}}& \textcolor{red}{\ding{55}}& 6.6 & 0.46 & 0.55 \\
     6&\textcolor{green}{\ding{51}}& \textcolor{green}{\ding{51}}& \textcolor{green}{\ding{51}}& \textbf{5.0}& \textbf{0.48} & \textbf{0.58} \\
        \bottomrule
    % \vspace{-0.9cm}
    \end{tabular}}}
    
    \label{tab:ablation}
    \vspace{-0.5cm}
\end{table}

\begin{table*}[!h]
    \centering
    %\huge
    \caption{Zero-shot TTS performance comparison between MELA-TTS and results from literature on seed-tts-eval. $\dag$ indicates that the model is trained using the same data, so the results are comparable.}
    % SS1 is calculated using WavLM-based speaker verification model and SS2 is calculated using ERes2Net-based speaker verification model.
    \resizebox{0.8\textwidth}{!}{
    {
    \begin{tabular}{lccccccccc}
        \toprule
			\multirow{2}{*}{\textbf{Model}} & \multicolumn{3}{c}{\textbf{\emph{test-zh}}} & \multicolumn{3}{c}{\textbf{\emph{test-en}}} & \multicolumn{3}{c}{\textbf{\emph{test-hard}}}\\
			\cmidrule(r){2-4} \cmidrule(r){5-7} \cmidrule(r){8-10}
			& \textbf{CER~$\downarrow$} & \multicolumn{1}{c}{\textbf{SS1~$\uparrow$}} & \multicolumn{1}{c}{\textbf{SS2~$\uparrow$}} & \textbf{WER~$\downarrow$} & \multicolumn{1}{c}{\textbf{SS1~$\uparrow$}} & \multicolumn{1}{c}{\textbf{SS2~$\uparrow$}}  & \textbf{CER~$\downarrow$} & \multicolumn{1}{c}{\textbf{SS1~$\uparrow$}} & \multicolumn{1}{c}{\textbf{SS2~$\uparrow$}} \\
			\midrule
			\textbf{Human} & 1.3 & 0.76 & 0.78 & 2.1 & 0.73 & 0.74  & - & -\\
            \midrule
            \multicolumn{10}{c}{\textbf{Non-autoregressive Models}} \\
             \midrule
            F5-TTS~\cite{chen-etal-2024-f5tts} & 1.6 & 0.74 & 0.80 & 1.8 & 0.65&0.74 & 8.7 & 0.71&0.76\\
            MaskGCT~\cite{maskgct} & 2.3 & 0.77 & 0.75& 2.6 &  0.71 & 0.73 & 10.3 & 0.75 & 0.72\\
            \midrule
            \multicolumn{10}{c}{\textbf{Autoregressive Models}} \\
             \midrule
            Seed-TTS~\cite{seed-tts} & 1.1 & \textbf{0.80} & - & 2.3 & \textbf{0.76} & - & 7.6 & \textbf{0.78} & -\\
            DiTAR~\cite{ditar} & 1.0 & 0.75 & - & \textbf{1.7} & 0.74 & - & - & - & -\\
            CosyVoice~\cite{cosyvoice} $\dag$  & 3.6 & 0.72&0.78 & 4.3 & 0.61&0.70 & 11.8 & 0.71&0.76\\
            CosyVoice 2.0~\cite{cosyvoice2} $\dag$  & 1.5 & 0.75&0.81 & 2.6 & 0.65&0.74 & \textbf{6.8} & 0.72&0.78\\
            CosyVoice 3.0-0.5B~\cite{cosyvoice3} $\dag$ & 1.3 & 0.75&\textbf{0.81} & 2.5 & 0.65&\textbf{0.75} & 7.0 & 0.72&\textbf{0.79}\\
            \midrule
            \multicolumn{10}{c}{\textbf{MELA-TTS}} \\
             \midrule
            w/o rep align $\dag$ & 1.2 & 0.74&0.79 & 4.0 & 0.60&0.68 & 10.9 & 0.72&0.78 \\
            w/ rep align $\dag$ & \textbf{0.9} & 0.72&0.77 & 2.4 & 0.59&0.68 & 7.6 & 0.71&0.76\\
            streaming mode w/ rep align $\dag$ & \textbf{0.9} & 0.72 & 0.78 & 2.5 & 0.59 & 0.68 & 7.7 & 0.71 & 0.77 \\
        \bottomrule
    \end{tabular}}}
    \vspace{-0.5cm}
    \label{tab:main}
\end{table*}
\section{Experiments}
\label{exp}
\subsection{Experiment settings}
\label{sec:setting}

\subsubsection{Datasets}
The experiments are conducted on 585-hour LibriTTS~\cite{zen2019libritts}, and an in-house 170,000-hour dataset, including 130,000-hour Chinese, 30,000-hour English, and 10,000-hour other languages. Experiments on LibriTTS are mainly for ablation studies, and the dataset  of 170,000 hours is used to evaluate the scaling ability.

\subsubsection{Model configuration}
In MELA-TTS, waveforms are resampled at 24 kHz, and the feature is 80-dimensional mel-spectrogram extracted at 50 Hz with a window length of 1920 and a hop length of 480. The size of the mel chunk is 8 (160ms). Thus, the autoregressive transformer works at a rate of 6.25 Hz (50/8 Hz) to generate continuous vectors $\bf{h}$, much smaller than most discrete-token-based TTS systems (25Hz for CosyVoice3 and 75Hz for VALL-E). With the interleaving ratio $n:m=$ 4:3, MELA-TTS generates 3 mel-spectrogram chunks for every 4 text tokens. A HiFTNet-based vocoder~\cite{li2023hiftnet} is used to reconstruct the waveform from the mel-spectrograms.

The transformer decoder follows the configuration of the pre-trained textual LLM, Qwen2-0.5B~\cite{Qwen2}, and is initialized using its weights. The diffusion module is a 22-layer diffusion transformer with 1024-dimensional hidden states and 16 heads. Following~\cite{le2023voiceboxtextguidedmultilingualuniversal}, the diffusion module is trained on both conditional and non-conditional situations to enable the classifier-free guidance (CFG)~\cite{cfg} at inference:
$${\hat{\bf X}}^{(i)}_{0, {\rm cfg}} = (1+\alpha) {\rm DiT}(\Psi_i,[{\bf X}^{(i-1)}_0, {\bf X}^{(i)}_t]) - \alpha {\rm DiT}(\emptyset,[{\bf X}^{(i-1)}_0, {\bf X}^{(i)}_t]),$$
and $\alpha$ is set to 0.7. For sampling, we use the DDIM sampler~\cite{ddim}, which accelerates generation by adopting a deterministic sampling process, and the default number of function evaluations (NFE) is 10. 

We adopt the encoder of SenseVoice-Large~\cite{sensevoice} to produce semantic representations for representation alignment. Note that the input feature of the pre-trained ASR encoder is not necessarily the same as the feature we adopted for TTS. For SenseVoice-Large, the input waveform is resampled at 16 kHz, and a 128-dimensional mel-spectrogram is computed with a window length of 400 and a hop length of 160. The encoder downsample the mel-spectrogram by a factor of 4, yielding an output representation $\bf{h_{asr}}$ at 25Hz.  Thus, the time alignment module (TAM) upsample $\bf{h}$ by a factor of 4 to match the temporal resolution of $\bf{h_{asr}}$. 

\subsubsection{Metrics}
We evaluate MELA-TTS using CER/WER for content consistency, and cosine similarity between generated speech and reference speech for voice cloning speaker similarity~(SS). Specifically, we use Whisper-large V3~\cite{systran_fwhisper_large_v3} to calculate English WER and Paraformer~\cite{DBLP:conf/interspeech/GaoZ0Y22} for Chinese CER. SS is calculated on the speaker embedding extracted by WavLM-TDNN~\cite{chen2022wavlm}~(the result is denoted as SS1), or the ERes2Net speaker verification model~\cite{chen2023enhanced}~(denoted as SS2). For subjective evaluation, we conducted an additional A/B preference test with 15 participants, each of whom was asked to complete 15 paired comparisons of speech samples.

\subsection{Ablation study on LibriTTS}
\label{sec:ablation}
% To evaluate the effectiveness of MELA-TTS, a series of ablation studies is conducted on Seed-EN. 

We conducted ablation studies on LibriTTS and evaluated on seed-tts-eval test-en\cite{seed-tts} to quantify the individual and combined contribution of utterance embedding and representation alignment. As presented in Table~\ref{tab:ablation}, in offline mode, the model without either component establishes a baseline of WER = 6.3, SS1 = 0.46, and SS2 = 0.55. Incorporating representation alignment alone yields a substantial 1.0-point reduction in WER (6.3 $\rightarrow$ 5.3). Moreover, as shown in Figure~\ref{fig:wer_convergence}, representation alignment accelerates training by over 3.3$\times$, reaching comparable performance of the model trained over 100 epochs without representation alignment, in less than 30 epochs. Speaker similarity (SS1 0.46 $\rightarrow$ 0.47, SS2 0.55 $\rightarrow$ 0.57) improves when utterance embedding is introduced, which demonstrates its capacity to enhance the modeling ability of speaker information. Most significantly, the system that combines utterance embedding and representation alignment achieves the optimal offline performance with a WER of 5.2, SS1 of 0.48, and SS2 of 0.58, which shows clear synergy. We attribute this to the complementary roles of the two modules: representation alignment explicitly regularizes cross-modal semantic consistency, thereby freeing utterance embedding to specialize in fine-grained acoustic modeling, such as the speaker information. These findings underscore that jointly modeling utterance-level information and cross-modal alignment achieves a superior balance between content consistency and speaker similarity.

Directly using mel-spectrograms as the alignment target degrades both WER, SS1, and SS2 (Exp 2 vs. Exp 0), which suggests that aligning to pretrained ASR encoder representations is a more effective objective, presumably because it encourages a decoupled semantic-acoustic modeling: the autoregressive transformer produces semantically informative representations, and the acoustic details are reconstructed by the diffusion model. This is consistent with the findings in discrete-token-based TTS systems, where a decoupled semantic-acoustic modeling is proven to benefit both content consistency and voice cloning capability~\cite{cosyvoice, cosyvoice2}.

Table~\ref{tab:ablation} compares the performance of streaming and offline synthesis. In streaming mode, MELA-TTS exhibits comparable WER to offline mode: 6.3 vs. 6.6 for the baseline condition, and 5.2 vs. 5.0 when representation alignment and utterance embedding modules are incorporated. Furthermore, both SS1 and SS2 remain nearly identical to those obtained under the offline configuration, which demonstrates great robustness of MELA-TTS in streaming mode.

\subsection{Evaluation on large-scale data}
\label{sec:main results}
Results on 170,000-hour data are presented in Table~\ref{tab:main}. For all experiments on 170,000-hour data, utterance embedding is adopted by default.
Consistent with the findings on LibriTTS, the representation alignment module produces significant improvement on content consistency (25\%, 40\%, and 30\% relative CER/WER reduction on test-zh, test-en, and test-hard, respectively), with little degradation on speaker similarity. Streaming synthesis performs equally well as the offline mode. Compared with the results on LibriTTS, data scaling yields substantial performance improvement (WER  5.2 $\rightarrow$ 2.4, SS1 0.48 $\rightarrow$ 0.59 and SS2 0.58 $\rightarrow$ 0.68 on test-en), which demonstrates the superior scaling ability of MELA-TTS.

When compared to other recently proposed models, MELA-TTS with representation alignment achieves state-of-the-art WER/CER results on test-zh, much better than the discrete-token based counterpart CosyVoice using the same training data, and is comparable with other competitive models on test-en and test-hard. Notably, while continuous representation-based DiTAR achieves the lowest WER on test-en, it's not been evaluated on test-hard, so it's not clear whether it's robust enough on hard cases, e.g. generating long utterances with challenging patterns for autoregressive models, such as word repetitions, tongue twisters, and so on. 

For voice cloning, MELA-TTS is comparable with competitive models on tesh-zh and test-hard, but lags in test-en. A possible reason for the suboptimal performance on speaker similarity is that in MELA-TTS, the diffusion module can only leverage the local context, while in discrete-token-based multi-stage system, like CosyVoice series, the diffusion or flow-matching module can utilize all input tokens (all history tokens in the streaming mode), as well as the prompt speech, as conditions to generate the mel-spectrogram. Similar speaker similarity gaps have also been observed in other continuous-representation-based systems~\cite{melle, clear}. We leave the optimization of the voice cloning ability for future work.

\begin{figure}[!t]
  \centering
  \centerline{\includegraphics[width=7cm]{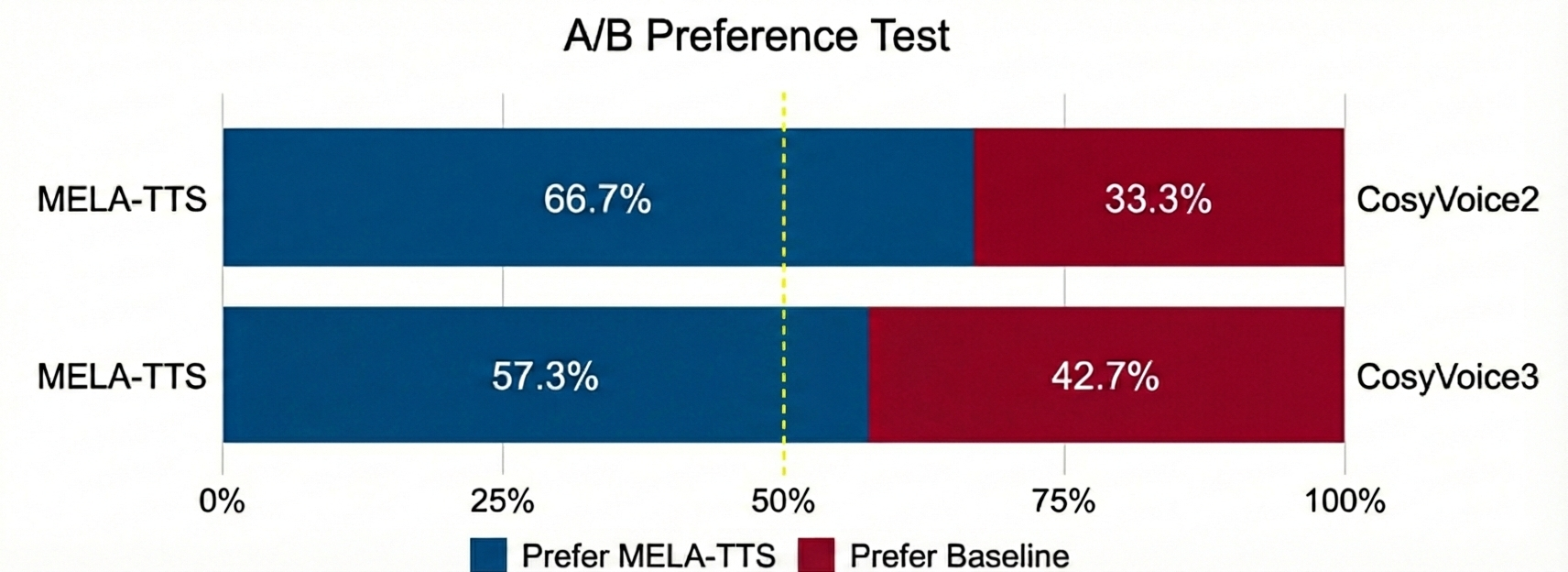}}
  \vspace{-0.5cm}
%  \vspace{2.0cm}
  % \centerline{}\medskip
\caption{Subjective preference between MELA-TTS and CosyVoice.}
\label{fig:preference_test}
\vspace{-0.6cm}
\end{figure}

The results of subjective evaluation are shown in Figure~\ref{fig:preference_test}. The proposed method is preferred over CosyVoice2 and CosyVoice3 in 66.7\% and 57.3\% of the trials, indicating better perceptual quality than discrete-token based counterparts.

\section{Conclusions}
\label{conclu}
We propose MELA-TTS, a joint transformer-diffusion framework for end-to-end text-to-speech synthesis, eliminating the dependency on speech tokenization and multi-stage processing pipelines. We further propose a representation alignment module to enhance the model's ability to capture semantic information. The proposed model is evaluated in both non-streaming and streaming modes, on datasets with scales varying from 585 to over 170,000 hours, demonstrating its effectiveness. In the future, we will further enhance the voice cloning capability of MELA-TTS and explore its applications in other domains, such as audio and music generation.

\vfill\pagebreak

\bibliographystyle{IEEEbib}
\bibliography{mela}

@article{valle,
  title={Neural codec language models are zero-shot text to speech synthesizers},
  author={Wang, Chengyi and Chen, Sanyuan and Wu, Yu and Zhang, Ziqiang and Zhou, Long and Liu, Shujie and Chen, Zhuo and Liu, Yanqing and Wang, Huaming and Li, Jinyu and others},
  journal={arXiv preprint arXiv:2301.02111},
  year={2023}
}

@article{cosyvoice3,
  title={Cosyvoice 3: Towards in-the-wild speech generation via scaling-up and post-training},
  author={Du, Zhihao and Gao, Changfeng and Wang, Yuxuan and Yu, Fan and Zhao, Tianyu and Wang, Hao and Lv, Xiang and Wang, Hui and Ni, Chongjia and Shi, Xian and others},
  journal={arXiv preprint arXiv:2505.17589},
  year={2025}
}

@article{cosyvoice2,
  title={Cosyvoice 2: Scalable streaming speech synthesis with large language models},
  author={Du, Zhihao and Wang, Yuxuan and Chen, Qian and Shi, Xian and Lv, Xiang and Zhao, Tianyu and Gao, Zhifu and Yang, Yexin and Gao, Changfeng and Wang, Hui and others},
  journal={arXiv preprint arXiv:2412.10117},
  year={2024}
}

@article{Qwen2,
  title={Qwen2 Technical Report},
  author={Qwen team},
  journal={arXiv preprint arXiv:2407.10671},
  year={2024},
}

@article{kaiming_ardiff,
  title={Autoregressive image generation without vector quantization},
  author={Li, Tianhong and Tian, Yonglong and Li, He and Deng, Mingyang and He, Kaiming},
  journal={Advances in Neural Information Processing Systems},
  volume={37},
  pages={56424--56445},
  year={2024}
}

@article{song2020score,
  title={Score-based generative modeling through stochastic differential equations},
  author={Song, Yang and Sohl-Dickstein, Jascha and Kingma, Diederik P and Kumar, Abhishek and Ermon, Stefano and Poole, Ben},
  journal={arXiv preprint arXiv:2011.13456},
  year={2020}
}

@article{ddim,
  title={Denoising diffusion implicit models},
  author={Song, Jiaming and Meng, Chenlin and Ermon, Stefano},
  journal={arXiv preprint arXiv:2010.02502},
  year={2020}
}

@article{dit,
  title={Scalable Diffusion Models with Transformers},
  author={William Peebles and Saining Xie},
  year={2022},
  journal={arXiv preprint arXiv:2212.09748},
}

@article{cosyvoice,
  title={Cosyvoice: A scalable multilingual zero-shot text-to-speech synthesizer based on supervised semantic tokens},
  author={Du, Zhihao and Chen, Qian and Zhang, Shiliang and Hu, Kai and Lu, Heng and Yang, Yexin and Hu, Hangrui and Zheng, Siqi and Gu, Yue and Ma, Ziyang and others},
  journal={arXiv preprint arXiv:2407.05407},
  year={2024}
}

@article{zen2019libritts,
  title={Libritts: A corpus derived from librispeech for text-to-speech},
  author={Zen, Heiga and Dang, Viet and Clark, Rob and Zhang, Yu and Weiss, Ron J and Jia, Ye and Chen, Zhifeng and Wu, Yonghui},
  journal={arXiv preprint arXiv:1904.02882},
  year={2019}
}

@misc{systran_fwhisper_large_v3,
	author       = { Systran },
	title        = { Faster Whisper Large V3},
	year         = 2023,
	url          = { https://huggingface.co/Systran/faster-whisper-large-v3 },
	publisher    = { Hugging Face }
}

@inproceedings{DBLP:conf/interspeech/GaoZ0Y22,
	author       = {Zhifu Gao and
	Shiliang Zhang and
	Ian McLoughlin and
	Zhijie Yan},
	title        = {Paraformer: Fast and Accurate Parallel Transformer for Non-autoregressive
	End-to-End Speech Recognition},
	booktitle    = {{Interspeech}},
	pages        = {2063--2067},
	publisher    = {{ISCA}},
	year         = {2022}
}

@article{chen2023enhanced,
  title={An enhanced res2net with local and global feature fusion for speaker verification},
  author={Chen, Yafeng and Zheng, Siqi and Wang, Hui and Cheng, Luyao and Chen, Qian and Qi, Jiajun},
  journal={arXiv preprint arXiv:2305.12838},
  year={2023}
}

@article{chen-etal-2024-f5tts,
      title={F5-TTS: A Fairytaler that Fakes Fluent and Faithful Speech with Flow Matching}, 
      author={Yushen Chen and Zhikang Niu and Ziyang Ma and Keqi Deng and Chunhui Wang and Jian Zhao and Kai Yu and Xie Chen},
      journal={arXiv preprint arXiv:2410.06885},
      year={2024},
}

@article{seed-tts,
      title={Seed-TTS: A Family of High-Quality Versatile Speech Generation Models}, 
      author={Philip Anastassiou and Jiawei Chen and Jitong Chen and Yuanzhe Chen and Zhuo Chen and Ziyi Chen and Jian Cong and Lelai Deng and Chuang Ding and Lu Gao and Mingqing Gong and Peisong Huang and Qingqing Huang and Zhiying Huang and Yuanyuan Huo and Dongya Jia and Chumin Li and Feiya Li and Hui Li and Jiaxin Li and Xiaoyang Li and Xingxing Li and Lin Liu and Shouda Liu and Sichao Liu and Xudong Liu and Yuchen Liu and Zhengxi Liu and Lu Lu and Junjie Pan and Xin Wang and Yuping Wang and Yuxuan Wang and Zhen Wei and Jian Wu and Chao Yao and Yifeng Yang and Yuanhao Yi and Junteng Zhang and Qidi Zhang and Shuo Zhang and Wenjie Zhang and Yang Zhang and Zilin Zhao and Dejian Zhong and Xiaobin Zhuang},
      year={2024},
      journal={arXiv preprint arXiv:2406.02430},
}

@article{sensevoice,
  author       = {Keyu An and
                  Qian Chen and
                  Chong Deng and
                  Zhihao Du and
                  Changfeng Gao and
                  Zhifu Gao and
                  Yue Gu and
                  Ting He and
                  Hangrui Hu and
                  Kai Hu and
                  Shengpeng Ji and
                  Yabin Li and
                  Zerui Li and
                  Heng Lu and
                  Haoneng Luo and
                  Xiang Lv and
                  Bin Ma and
                  Ziyang Ma and
                  Chongjia Ni and
                  Changhe Song and
                  Jiaqi Shi and
                  Xian Shi and
                  Hao Wang and
                  Wen Wang and
                  Yuxuan Wang and
                  Zhangyu Xiao and
                  Zhijie Yan and
                  Yexin Yang and
                  Bin Zhang and
                  Qinglin Zhang and
                  Shiliang Zhang and
                  Nan Zhao and
                  Siqi Zheng},
  title        = {FunAudioLLM: Voice Understanding and Generation Foundation Models for Natural Interaction Between Humans and LLMs},
  journal      = {arXiv preprint arXiv:2407.04051},
  year         = {2024},

}

@article{maskgct,
	author       = {Yuancheng Wang and
	Haoyue Zhan and
	Liwei Liu and
	Ruihong Zeng and
	Haotian Guo and
	Jiachen Zheng and
	Qiang Zhang and
	Shunsi Zhang and
	Zhizheng Wu},
	title        = {MaskGCT: Zero-Shot Text-to-Speech with Masked Generative Codec Transformer},
	journal      = {arXiv preprint arXiv:2409.00750},
year={2024},
}

@misc{ditar,
      title={DiTAR: Diffusion Transformer Autoregressive Modeling for Speech Generation}, 
      author={Dongya Jia and Zhuo Chen and Jiawei Chen and Chenpeng Du and Jian Wu and Jian Cong and Xiaobin Zhuang and Chumin Li and Zhen Wei and Yuping Wang and Yuxuan Wang},
      year={2025},
journal      = {arXiv preprint arXiv:2502.03930},

}

@misc{le2023voiceboxtextguidedmultilingualuniversal,
      title={Voicebox: Text-Guided Multilingual Universal Speech Generation at Scale}, 
      author={Matthew Le and Apoorv Vyas and Bowen Shi and Brian Karrer and Leda Sari and Rashel Moritz and Mary Williamson and Vimal Manohar and Yossi Adi and Jay Mahadeokar and Wei-Ning Hsu},
      year={2023},
    journal      = {arXiv preprint arXiv:2306.15687},
}

@misc{melle,
      title={Autoregressive Speech Synthesis without Vector Quantization}, 
      author={Lingwei Meng and Long Zhou and Shujie Liu and Sanyuan Chen and Bing Han and Shujie Hu and Yanqing Liu and Jinyu Li and Sheng Zhao and Xixin Wu and Helen Meng and Furu Wei},
      year={2025},
 journal      = {arXiv preprint arXiv:2407.08551},

}

@article{chen2022wavlm,
  title={Wavlm: Large-scale self-supervised pre-training for full stack speech processing},
  author={Chen, Sanyuan and Wang, Chengyi and Chen, Zhengyang and Wu, Yu and Liu, Shujie and Chen, Zhuo and Li, Jinyu and Kanda, Naoyuki and Yoshioka, Takuya and Xiao, Xiong and others},
  journal={IEEE Journal of Selected Topics in Signal Processing},
  volume={16},
  number={6},
  pages={1505--1518},
  year={2022},
  publisher={IEEE}
}

@article{cfg,
  title={Classifier-free diffusion guidance},
  author={Ho, Jonathan and Salimans, Tim},
  journal={arXiv preprint arXiv:2207.12598},
  year={2022}
}

@article{kalle,
      title={Autoregressive Speech Synthesis with Next-Distribution Prediction}, 
      author={Xinfa Zhu and Wenjie Tian and Lei Xie},
      year={2024},
journal={arXiv preprint arXiv:2412.16846},

}

@article{clear,
      title={CLEAR: Continuous Latent Autoregressive Modeling for High-quality and Low-latency Speech Synthesis}, 
      author={Chun Yat Wu and Jiajun Deng and Guinan Li and Qiuqiang Kong and Simon Lui},
journal={arXiv preprint arXiv:2508.19098},
      year={2025},
}

@article{li2023hiftnet,
  title={Hiftnet: A fast high-quality neural vocoder with harmonic-plus-noise filter and inverse short time fourier transform},
  author={Li, Yinghao Aaron and Han, Cong and Jiang, Xilin and Mesgarani, Nima},
  journal={arXiv preprint arXiv:2309.09493},
  year={2023}
}

\end{document}